\documentstyle[preprint,aps,prb,epsfig]{revtex}
\begin{document}
\draft
\pagestyle{plain}
\newcommand{\D}{\displaystyle}
\title{\bf Impurity scattering effect on the specific heat jump in anisotropic 
superconductors.}
\author{Grzegorz Hara\'n\cite{AA}, Jason Taylor and A. D. S. Nagi}
\vspace{0.4cm}
\address{Department of Physics,
University of Waterloo,
Waterloo, Ontario,
Canada, N2L 3G1}
\date{December 31, 1996} 
\maketitle

\begin{abstract}
The specific heat jump at a normal-superconducting phase transition 
in an anisotropic superconductor with nonmagnetic impurities is calculated 
within a weak-coupling mean field approximation. 
It is shown that its dependence on the impurity concentration is 
remarkably different for $d_{x^2-y^2}$-wave and ($d_{x^2-y^2}+s$)-wave states.   
This effect may be used as a test for the presence of an $s$-wave 
admixture in the cuprates. 
\end{abstract}

There now exists a considerable experimental evidence supporting $d$-wave 
superconductivity in the cuprates, but the most direct probes of the   
superconducting state like the electromagnetic penetration depth, photoemission     
and quantum phase interference measurements neither confirm nor exclude  
a possible small $s$-wave admixture in a predominantly $d_{x^2-y^2}$ 
superconductor. \cite{1} The linear temperature dependence of the   
penetration depth at low temperatures \cite{17}  
observed in YBCO agrees with the theoretical predictions for 
a $d$-wave state. \cite{18} However, the measurements only went down to 
about 1K and an exponential behavior below this temperature, indicating  
a small nonzero gap minimum cannot be eliminated.  
Even by taking data at much lower temperatures, the presence of a small 
$s$-wave component in the order parameter cannot be entirely excluded 
in the penetration depth experiments. Similar constraints limit the 
angle resolved photoemission spectroscopy (ARPES) method. Although ARPES  
data \cite{1,19,20} are consistent with a $d_{x^2-y^2}$ scenario 
in BSCCO as well as in YBCO, the experiments cannot 
decide with an accuracy greater than an instrumental resolution  
if the order parameter completely vanishes at the $d_{x^2-y^2}$ nodal lines.  
This leaves the possibility of a small ($<$ 2meV) $s$-wave admixture. Therefore  
the above experimental methods do not rule out the presence of a 
small isotropic component, but place an upper bound on the minimum of the 
gap function. As analyzed in Ref. 1, the emerging picture from the Josephson 
experiments supports a scenario of a real mixture of $s$ and $d_{x^2-y^2}$ 
states in YBCO, but also does not definitely confirm the presence of the     
$s$-wave component. The existence of even a small $s$-wave admixture in  
a $d$-wave superconductor may be tested by thermodynamic measurements 
in the presence of nonmagnetic impurities. 
It is well known that the $d$-wave state is strongly suppressed by the defects,  
\cite{9,16} and the $s$-wave state is not affected by the nonmagnetic scatterers.  
\cite{8} In the case of ($d+s$)-wave superconductor a power-law $T_c$   
suppression \cite{7} should be observed above certain impurity-doping level 
and the thermodynamic properties at large impurity concentration should 
resemble those of the $s$-wave state. 
In fact the critical temperature of YBCO is decreased below 12K by the 
electron-irradiation, \cite{1} and the Pr-doping or ion-beam damage lead 
to a long tail $T_c$ suppression \cite{11} characteristic for a small 
nonzero value of the gap function integrated over the Fermi surface. \cite{7}  
However, despite the electron-irradiation removing the planar oxygens produces 
the nonmagnetic defects, \cite{21} it has not been determined, whether 
the scattering centers created by Pr-doping and ion-beam damage in YBCO are 
purely nonmagnetic.

In the present paper we suggest that more significant features attributed 
to the $s$-wave part of the order parameter may be seen in the specific  
heat measurements. We study a nonmagnetic impurity effect on the  
specific heat jump at a superconducting-normal phase transition  
in anisotropic superconductors and show that the result depends 
on the Fermi surface (FS) averages of the first four powers of the 
superconducting order parameter. A particularly large difference 
in the specific heat jump between the states with a nonzero and a zero value of 
the order parameter FS average is observed. We suggest that  
this measurement may be used as a test for the presence of an $s$-wave 
admixture in a $d_{x^2-y^2}$ state. We take $\hbar=k_B=1$ throughout 
the paper.\\

We consider the effect of potential scattering by nonmagnetic, 
noninteracting impurities on the order parameter with its orbital 
part defined as follows 

\begin{equation}
\label{e1}
\Delta\left({\bf k}\right)=\Delta e\left({\bf k}\right)
\end{equation}

\noindent
where $e\left({\bf k}\right)$ is a momentum-dependent function  
normalized by taking its average value over the Fermi surface 
$\left<e^{2}\right>=\int_{FS}dS_k n\left({\bf k}\right)e^2\left({\bf k}\right)=1$,  
where $\int_{FS}dS_{k}$ represents the integration over the Fermi surface and  
$n\left({\bf k}\right)$ is the angle resolved FS density of states,  
which obeys $\int_{FS}dS_k n\left({\bf k}\right)=1$. 
This normalization gives $\Delta$  
the meaning of the absolute magnitude of the order parameter. 
The function $e\left({\bf k}\right)$ may belong to a one 
dimensional (1D) irreducible representation of the crystal point 
group or may be given by a linear combination of the basis functions 
of different 1D representations. The impurity effect is studied  
in the t-matrix approximation. \cite{28,2} 
This approach introduces two parameters describing the scattering 
process: $c=1/(\pi N_0 V_i)$ and $\Gamma=n_i/\pi N_0$, where 
$N_{0}$, $V_i$ and $n_i$ are respectively the overall density of  
states at the Fermi level, the impurity (defect) potential 
and the impurity concentration. We assume an $s$-wave scattering 
by the impurities, that is $V_i$ does not have an internal momentum-
dependence. It is particularly convenient to think of $c$ as a measure of the  
scattering strength, with $c\rightarrow 0$ in the unitary limit and $c\gg 1$ for 
weak scattering i.e. the Born limit. 

The amplitude of the order parameter is determined by the mean-field 
self-consistent equation   

\begin{equation}
\label{e7}
\Delta\left({\bf k}\right)=-T\sum_{\omega}\sum_{{\bf k'}}
V\left({\bf k},{\bf k'}\right)
\frac{\tilde{\Delta}\left({\bf k'}\right)}
{\tilde{\omega}^{2}+{\xi_{k'}}^{2}+|\tilde{\Delta}\left(
{\bf k'}\right)|^{2}}
\end{equation}

\noindent
where T is the temperature, $\xi_{k}$ is the quasiparticle energy,  
$\omega=\pi T(2n+1)$ (n is an integer), and $V\left({\bf k}, {\bf k'}\right)$  
is the phenomenological pair potential taken as 

\begin{equation}
\label{e8}
V\left({\bf k}, {\bf k'}\right)=-V_{0}e\left({\bf k}\right)
e\left({\bf k'}\right)
\end{equation}

\noindent
We have assumed a particle-hole symmetry of a quasiparticle spectrum. 
The renormalized Matsubara frequency $\tilde{\omega}\left({\bf k}\right)$ 
and the renormalized order parameter $\tilde{\Delta}\left({\bf k}\right)$  
are then given by 

\begin{eqnarray}
\label{e6}
\tilde{\omega}=\omega-\Sigma_0,&\;\;\;\;&
\tilde{\Delta}\left({\bf k}\right)=\Delta\left({\bf k}\right)+\Sigma_1
\end{eqnarray}

\noindent
with the self-energies defined as follows 

\begin{eqnarray}
\label{e6a}
\D\Sigma_0=-\Gamma\frac{g_0}{c^2+g^{2}_{0}+g^{2}_{1}},&\;\:\;\;&
\D\Sigma_1=\Gamma\frac{g_1}{c^2+g^{2}_{0}+g^{2}_{1}}
\end{eqnarray}

\noindent
and the $g_0$, $g_1$ functions determined by the self-consistent equations 

\begin{equation}
\label{e6b}
\D g_0=\frac{1}{N_0\pi}\sum_{\bf k}\frac{\tilde{\omega}}
{\tilde{\omega}^{2}+{\xi_{k}}^{2}+|\tilde{\Delta}\left({\bf k}\right)|^{2}}
\end{equation}

\begin{equation}
\label{e6c}
\D g_1=\frac{1}{N_0\pi}\sum_{\bf k}\frac{\tilde{\Delta}\left({\bf k}\right)} 
{\tilde{\omega}^{2}+{\xi_{k}}^{2}+|\tilde{\Delta}\left({\bf k}\right)|^{2}}
\end{equation}

\noindent
To proceed further, we restrict the wave vectors of the electron self-energy 
and pairing potential to the Fermi surface and replace $\sum_{\bf k}$  
by $N_{0}\int_{FS}dS_{k}n\left({\bf k}\right)\int d\xi_{k}$. 
Integrated over $\xi_{k}$ the gap equation (\ref{e7}) can be transformed 
after a standard procedure \cite{13} into 

\begin{equation}
\label{e8a}
\ln\left(\frac{T}{T_{c_{0}}}\right)=2\pi T\sum_{\omega>0}
\left(f_{\omega}-\frac{1}{\omega}\right)
\end{equation}

\noindent
where the $f_{\omega}$ function is defined as follows

\begin{equation}
\label{e8b}
f_{\omega}=\D\int_{FS}dS_{k}n\left({\bf k}\right)
\frac{\tilde{\Delta}\left({\bf k}\right)e\left({\bf k}\right)}
{\Delta\left[\tilde{\omega}^{2}+|\tilde{\Delta}\left(
{\bf k}\right)|^{2}\right]^{\frac{1}{2}}}
\end{equation}

\noindent
We expand Eq. (\ref{e8a}) in powers of $\Delta^{2}$ around 
$\Delta=0$ using the relations (\ref{e6})-(\ref{e6c}). 
Keeping up to the fourth power terms in $\Delta$ we get the gap equation   
in the Ginzburg-Landau regime  

\begin{equation}
\label{e9}
\ln\left(\frac{T}{T_{c_{0}}}\right)=-f_{0}-\frac{1}{2}f_{1}
\left(\frac{\Delta}{2\pi T}\right)^{2}+
\frac{1}{4}f_{2}\left(\frac{\Delta}{2\pi T}\right)^{4}
\end{equation}

\noindent
where the coefficients are given by

\begin{equation}
\label{e10}
f_{0}=-2\pi T\sum_{\omega>0}\left(\left(f_{\omega}\right)_{\Delta=0}
-\frac{1}{\omega}\right)
\end{equation}

\begin{equation}
\label{e11}
f_{1}=-\left(2\pi T\right)^{3}\sum_{\omega}
\left(\frac{df_{\omega}}{d\Delta^{2}}\right)_{\Delta=0}
\end{equation}

\begin{equation}
\label{e12}
f_{2}=2\left(2\pi T\right)^{5}\sum_{\omega}
\left(\frac{d^{2}f_{\omega}}{d\left(\Delta^{2}\right)^{2}}
\right)_{\Delta=0}
\end{equation}

\noindent
Taking the derivatives with respect to $\Delta^2$ 

\begin{equation}
\label{e25}
\D\frac{d\;\;\;}{d\Delta^{2}}= \frac{\partial\;\;\;}{\partial\Delta^{2}}+
\sum_{\omega}\left\{\frac{d\tilde{\omega}}
{d\Delta^{2}}\frac{\partial\;\;}{\partial\tilde{\omega}}  
+\frac{d\tilde{\Delta}\left({\bf k}\right)}
{d\Delta^{2}}\frac{\partial\;\;\;\;\;\;}
{\partial\tilde{\Delta}\left({\bf k}\right)}\right\}   
\end{equation}

\noindent
and with a use of the relations given in Eqs. (\ref{e6})-(\ref{e6c}) we 
calculate $f_0$ and $f_1$ coefficients 

\begin{equation}
\label{e13}
\begin{array}{l}
\D f_{0}\left(\varrho\right)=\left(1-\left<e\right>^{2}\right)
\left(\psi\left(\frac{1}{2}+\varrho\right)
-\psi\left(\frac{1}{2}\right)\right)\;\;\;\;\;\;\;\;\;\;\;\;\;\;\; 
\;\;\;\;\;\;\;\;\;\;\;\;\;\;\;\;\;\;\;\;\;\;\;\;\;\;\;\;\;\;\;\;\;
\;\;\;\;\;\;\;\;\;\;\;\;\;
\end{array}
\end{equation}

\begin{equation}
\label{e14a}
\begin{array}{l}
\D f_{1}\left(\varrho\right)=2\left<e\right>\left[2\left<e^3\right>+5\left<e\right>^3 
-7\left<e\right>\right]\varrho^{-2}\left(\psi\left(\frac{1}{2}
+\varrho\right)-\psi\left(\frac{1}{2}\right)\right)\\
\\
\D +2\left<e\right>\left[-2\left<e^{3}\right>-3\left<e\right>^3 
+5\left<e\right>\right] 
\varrho^{-1}\psi^{(1)}\left(\frac{1}{2}+\varrho\right)
+4\left<e\right>^2\left[1-\left<e\right>^{2}\right]\varrho^{-1} 
\psi^{(1)}\left(\frac{1}{2}\right)\\
\\
\D +\frac{1}{2}\left[-\left<e^{4}\right>+3\left<e\right>^{4}
+4\left<e\right>\left<e^{3}\right>-6\left<e\right>^{2}\right] 
\psi^{(2)}\left(\frac{1}{2}+\varrho\right)
-\frac{1}{2}\left<e\right>^{4}\psi^{(2)}\left(\frac{1}{2}\right)\\
\\
\D +\frac{1}{6}\left[2\left(\left<e\right>^{2}-1\right)^2\frac{1}{c^2+1} 
-\left<e\right>^{4}+2\left<e\right>^{2}-1\right] 
\varrho\psi^{(3)}\left(\frac{1}{2}+\varrho\right)
\end{array}
\end{equation}

\noindent
where $\varrho=\left[\Gamma/(c^2+1)\right]/\left(2\pi T\right)$ and $\psi$,  
$\psi^{(n)}\;(n=1,2,3)$ are the polygamma functions. \cite{10} 
In the unitary limit $c\rightarrow 0$ and $\varrho=\Gamma/(2\pi T)$. Alternatively  
for weak scattering ($c\gg 1$) we keep only the terms linear in $1/c^2$ in  
a Taylor's expansion which leads to the Born approximation scattering rate    
$\varrho=\pi N_0 n_i V^2_i/(2\pi T)$ and $\varrho/(c^2+1)=0$.  
Coefficients $f_0$ and $f_1$ involve three different types of  
the Fermi surface averages of the order parameter namely, 
$\left<e\right>$, $\left<e^{3}\right>$, and $\left<e^{4}\right>$. These 
averages enter the free energy and determine the thermodynamic properties 
at the phase transition. In this paper we discuss a  specific heat jump at  
$T_c$, $\Delta C(T_c)=C_S(T_c)-C_N(T_c)$, where $C_S(T_c)$ and $C_N(T_c)$,    
respectively are the specific heat of the superconducting and normal state,  
$C_N(T_c)=(2\pi^2/3)N_0T_c$.         
We obtain \cite{13} from Eq. (\ref{e9}) that  

\begin{equation}
\label{e14}
\D\frac{\Delta C\left(T_c\right)}{C_N\left(T_c\right)}
=\frac{12}{\left(f_1\right)_{T=T_c}}
\left[1+T_c\left(\frac{df_0}{dT}\right)_{T=T_c}\right]^{2} 
\end{equation}

\noindent
and finally, $f_0$ from Eq. (\ref{e13}) yields  

\begin{equation}
\label{e15}
\D\frac{\Delta C\left(T_c\right)}{C_{N}\left(T_c\right)}
=\frac{12}{f_{1}\left(\varrho_{c}\right)}
\left[1+\left(\left<e\right>^2-1\right)\varrho_{c}\psi^{(1)}\left(\frac{1}{2}+
\varrho_{c}\right)\right]^{2}
\end{equation}

\noindent
where $\varrho_c$ is $\varrho$ at $T=T_c$. 
This rather cumbersome formula, when considered along with Eq. (\ref{e14a}),  
reduces significantly for $\left<e\right>=0$ case 

\begin{equation}
\label{e15a}
\D\frac{\Delta C\left(T_c\right)}{C_{N}\left(T_c\right)}
=\frac{\D 12\left[1-\varrho_c\psi^{(1)}
\left(\frac{1}{2}+\varrho_c\right)\right]^{2}}
{\D\frac{\mu}{6}\varrho\psi^{(3)}\left(\frac{1}{2}+\varrho_c\right)
-\frac{1}{2}\left<e^{4}\right>
\psi^{(2)}\left(\frac{1}{2}+\varrho_c\right)}
\end{equation}
  
\noindent
where $\mu=(1-c^2)/(1+c^2)$.  
For an appropriate choice of $\left<e^{4}\right>$ value,   
$\Delta C\left(T_c\right)/C_{N}\left(T_c\right)$ from Eq. (\ref{e15a}) agrees 
with the result obtained by Hirschfeld et al. \cite{2} as well as that by  
Suzumura and Schulz \cite{15} in the Born limit.   

It is informative to discuss the limiting cases of Eq. (\ref{e15}), 
that is a pure system where $\varrho_c=0$ and a highly impure one 
with $\varrho_c\rightarrow\infty$ in which $T_c\rightarrow 0$  
suppressed by the impurities. Using a series representation of 
$f_1$ function \cite{3} we get in $\varrho_c=0$ limit 

\begin{equation}
\label{e16}
\D\left(\frac{\Delta C\left(T_c\right)}{C_{N}\left(T_c\right)}\right)_{\varrho_c=0}
=-\frac{24}{\D\psi^{(2)}\left(\frac{1}{2}\right)\left<e^{4}\right>}
\approx\frac{1.426}{\left<e^{4}\right>}
\end{equation}

\noindent
The $\varrho_c\rightarrow\infty$ limit is obtained with a use of 
Eq. (\ref{e14a}) and asymptotic forms of polygamma functions. \cite{10}  
There are two cases to distinguish here. First, when the Fermi surface 
average of the order parameter $\left<e\right>\neq 0$ then 
  
\begin{equation}
\label{e17}
\D\left(\frac{\Delta C\left(T_c\right)}{C_{N}\left(T_c\right)}\right)
_{\varrho_c\rightarrow\infty}=
-\frac{24}{\D\psi^{(2)}\left(\frac{1}{2}\right)}\approx 1.426 
\end{equation}

\noindent
and the second, \cite{29} with $\left<e\right>=0$, which leads to  

\begin{equation}
\label{e18}
\D\left(\frac{\Delta C\left(T_c\right)}{C_{N}\left(T_c\right)}\right)
_{\varrho_c\rightarrow\infty}=0
\end{equation}
 
\noindent
We note, that a specific heat jump value in $\varrho_c\rightarrow\infty$ 
limit for a nonzero value of $\left<e\right>$ 
given by Eq. (\ref{e17}) agrees 
with that of an isotropic s-wave superconductor. This fact has a simple 
intuitive interpretation. A nonzero Fermi surface average of the 
order parameter leads to an asymptotic power-law critical temperature 
suppression for large impurity concentration \cite{7}   
$T_c\sim\left(T_{c_{0}}\right)^{1/\left<e\right>^2}\left[\Gamma/
\left(c^2+1\right)\right]^{(1-1/\left<e\right>^2)}$,  
therefore $T_c$ is almost constant for 
large $\Gamma$ values. The impurity effect, then, 
in the large impurity concentration range is the same as in the case 
of s-wave superconductivity, where $T_c$ is not changed by the 
nonmagnetic impurities. Indeed, as it has been shown for the representative order 
parameters, \cite{34,35} the gap anisotropy is smeared out by the isotropic 
scattering when $\left<e\right>\neq 0$ and the density of states  
approaches that of an isotropic $s$-wave superconductor.  
Alternatively, for $\left<e\right>=0$ we observe 
a strong impurity-induced suppression of the critical temperature 
\cite{9,16} leading to a  
zero value at finite impurity concentration, which is reflected by  
a zero specific heat jump limit value in Eq. (\ref{e18}).  
As a nonzero value of $\left<e\right>$ can be achieved only when 
$e\left({\bf k}\right)$ contains a component belonging to an identity 
representation of the crystal point group, the measurement of the specific 
heat jump at the phase transition in the limit of $T_c\rightarrow 0$ (and  
large impurity concentration) may be used as a stringent test for the 
occurrence of even a small $A_{1g}$ admixture to the order parameter.    
It should be noted that the effect at large impurity concentration 
for $\left<e\right>\neq 0$ (Eq. (\ref{e17})) is independent 
of the amount of the $s$-wave content in the order parameter, however, 
as we discuss below, it may be hard to detect for a very small 
$s$-wave component as it would require an experiment at low temperatures.\\  

We discuss our results in a context of high-$T_c$ superconductivity, 
considering a $d_{x^2-y^2}$ state, \cite{1} that is the order 
parameter given by Eq. (\ref{e1}) with 
$e\left({\bf k}\right)=\left(k_{x}^{2}-k_{y}^{2}\right)
\left<\left(k_{x}^{2}-k_{y}^{2}\right)^2\right>^{-1/2}$.  
As we have mentioned above, our main result that is the value of the 
specific heat jump at $T_c\rightarrow 0$ is independent of the amplitude 
of the $s$-wave component and its origin. However, in order to establish  
a quantitative behavior of $\Delta C(T_c)$ in a whole range of impurity-doping 
we must work with a particular level of $s$-wave admixture. We do this by assuming  
that the $s$-wave component is an artifact of an orthorhombic anisotropy of the 
system and relate the amount of the $s$-wave admixture to the degree of this 
anisotropy. \cite{7,23} This approach gives a semi-microscopic justification 
for the $(d+s)$-wave state.  
The orthorhombicity in the case of YBCO means that the $a-$  
and $b-$ crystal axes in the $CuO_{2}$ planes become inequivalent, which  
leads, with a simple approximation of an elliptical Fermi surface,  
to the following form of an energy band \cite{7}    

\begin{equation}
\label{e2}
\xi_{\bf k}=c_xk_{x}^{2}+c_yk_{y}^{2}-\varepsilon_{F}
\end{equation}

\noindent
where a ratio of the effective masses $c_x/c_y$ is a dimensionless 
parameter describing the orthorhombic anisotropy 
of the Fermi surface and $\varepsilon_{F}$ is the Fermi energy. It is easy  
to see within this model, that a ($d_{x^2-y^2}+s$) state emerges 
from a $d_{x^2-y^2}$ in a natural way due to the orthorhombic    
distortion of the crystal lattice. A straightforward calculation   
based on a transformation from an elliptical FS to 
a circular one shows that the normalized $d_{x^2-y^2}$ order 
parameter defined on the FS given by Eq. (\ref{e2}) 
can be represented on a circular FS as 

\begin{equation}
\label{e3}
\D\Delta\left({\bf k}\right)=\Delta\frac{\D 1+\frac{c_x}{c_y}}
{\D\left[\frac{3}{2}-\frac{c_x}{c_y}+\frac{3}{2}\left(\frac{c_x}{c_y}\right)^{2}
\right]^{\frac{1}{2}}}\left[cos2\varphi+\frac{\D 1-\frac{c_x}{c_y}}
{\D 1+\frac{c_x}{c_y}}\right]
\end{equation}

\noindent
where $\varphi$ is the polar angle. In order to clarify the terminology, 
we will refer to the circular Fermi surface when classifying the superconducting 
states. Therefore, as a $d_{x^2-y^2}$ we define a state with 
$e\left({\bf k}\right)$ proportional to $cos2\varphi$ and the states 
with a nonzero $s$-wave contribution are called ($d_{x^2-y^2}+s$). 
We note, that the order parameter from Eq. (\ref{e3}) is $d_{x^2-y^2}$ 
when $c_x/c_y=1$ only, that is for a tetragonal symmetry, otherwise it   
contains a nonzero $s$-wave component proportional to $(1-c_x/c_y)$.    
In Tab. I we present as the functions of the orthorhombic anisotropy  
parameter $c_x/c_y$ the Fermi surface averages which enter the Ginzburg-Landau 
coefficients $f_0$ and $f_1$ given in Eqs. (\ref{e13}) and (\ref{e14a}).  
We emphasize, that the assumption of the orthorhombic asymmetry as the 
mechanism producing the $s$-wave admixture in the order parameter 
does not affect the results since only the amplitude of this component matters  
in the calculation. Thus one can obtain the same results  
in a more phenomenological way assuming the presence of the $s$-wave 
phase and taking its level as given by $\left<e\right>$ in Tab. I for 
the $c_x/c_y$ values considered in this paper. \cite{27}     

Based on the discussion of the specific heat jump for a large impurity 
concentration in Eqs. (\ref{e17})-(\ref{e18}) we can discuss this 
limit for $d$- and $(d+s)$-wave superconductors. For a pure $d_{x^{2}-y^{2}}$  
state ($c_x/c_y=1$) we have $\left<e\right>=0$.  
Therefore the specific heat jump decreases to zero with a critical 
temperature driven to zero by impurities as in Eq. (\ref{e18}). 
On the other hand, for even a slight s-wave component, $\left<e\right>\neq 0$ 
and the specific heat jump increases and reaches a finite nonzero 
value at $T_{c}\rightarrow 0$ given by Eq. (\ref{e17}). Below, we present the  
specific heat jump at the phase transition normalized by the specific heat in 
a normal state as a function of the normalized impurity scattering rate 
$\varrho_c T_c/T_{c_{0}}$ in the Born limit (Fig. 1a), where 
$\varrho_c T_c/T_{c_{0}}=\pi N_0 n_i V^2_i/(2\pi T_{c_{0}})$ and in the 
unitary limit (Fig. 1b) with $\varrho_c T_c/T_{c_{0}}=\Gamma/(2\pi T_{c_{0}})$.  
Note that $N_0=(c_xc_y)^{-1/2}S/(2\pi\hbar^2)$, where $S$ is a sample surface 
area, hence $T_{c_{0}}$ is different for different values of $c_xc_y$ product. 
In the Figs. 2a and 2b we show the same $\Delta C(T_c)/C_N(T_c)$ data 
versus the normalized critical temperature $T_c/T_{c_{0}}$.  
The considered states contain a small $s$-wave admixture varying from about 
$8\%$ to $16\%$, therefore we observe a strong $T_c$ suppression by the 
nonmagnetic impurities and a fast decrease in the specific heat jump as 
long as a significant $d$-wave component is present. Once it is almost destroyed and  
the $s$-wave part, which is insensitive to the nonmagnetic defects, prevails,    
the BCS normalized specific heat jump value of about $1.426$ is restored  
in a sudden increase of $\Delta C(T_c)/C_N(T_c)$. The general tendency of the   
$T_c$ suppression, given by Eq. (\ref{e9}) at $\Delta=0$,   
changes at that doping level too and the critical temperature asymptotically 
goes to zero (Fig. 3).  For the sake of comparison 
we show in Fig. 4 the specific heat jump $\Delta C(T_c)$ normalized by $C_N(T_c)$   
as a function of the impurity scattering rate $\varrho_c T_c/T_{c_{0}}$ in the  
Born and unitary limits for the ($s+d_{x^2-y^2}$) state, where the $s$-wave 
component is large ($\sim 60\%$) and the $d_{x^2-y^2}$ part is considered as  
minor. 

One can notice from the above figures that the unitary and 
Born scattering limits differ for small values of the pair-breaking parameter 
and fall on the same curve in the range where practically the $s$-wave  
superconductivity is only left. The pair-breaking parameter $\varrho_cT_c/T_{c_{0}}$, 
however, has a different meaning in either case. 

We have mentioned before that a detection of a small $s$-wave component 
would need a measurement at low temperatures. For instance, in a superconductor 
of the critical temperature in a clean limit $T_{c_{0}}=90K$ the 
$s$-wave content of about $7.4\%$ ($\left<e\right>\simeq 0.074$) can be observed 
at a temperature of $\sim 2.5K$, which is the estimated position of 
$\Delta C(T_c)/C_N(T_c)$ minimum in Figs. 2a and 2b. This minimum is a place where  
a distinct signal from the $s$-wave component appears, therefore its position is of 
special interest for possible experiments. We have found the minimum coordinates 
$(\varrho_cT_c/T_{c_{0}})^{*}$ (Fig. 5) and $(T_c/T_{c_{0}})^{*}$ (Fig. 6) 
as the functions of the order parameter FS average value $\left<e\right>$, which 
multiplied by $100\%$ gives the $s$-wave fraction in per cent in the 
normalized to unity order parameter $\left<e^2\right>=1$. A plot of   
$(\varrho_cT_c/T_{c_{0}})^{*}$ vs $\left<e\right>$ in Fig. 5 may be also of the  
experimental use, since the scattering rate $\varrho_cT_c/T_{c_{0}}$  
is proportional to the impurity concentration, which can be estimated 
in the measurements. As one can see from Figs. 
5 and 6 the measurements at low temperatures are required for small $s$-wave 
admixtures, however, they are to be performed at the phase transition which 
should be accessible as long as $T_c$ is measurable. Assuming that a possible 
$s$-wave admixture is of the order of magnitude of the experimental resolution  
error ($\sim 2.5\;meV$) in the ARPES measurements \cite{20} of the smallest    
energy gap values, we can estimate its fraction as a ratio 
$2.5\;meV/34\;meV\simeq 0.07$, where $34\;meV$ is a measured maximum $|\Delta|$ 
value. Therefore from Fig. 6 we get that the abrupt rise in the normalized 
specific heat jump should be observed at $T_c\sim 2.5\;K$ in a superconductor  
of the critical temperature in the absence of impurities $T_{c_{0}}=90K$.  
Experiments \cite{24} investigating the disorder effect on the specific heat 
jump at $T_c$ in YBCO show $\Delta C(T_c)/T_c$ suppression to zero with the 
increasing impurity concentration. However, the magnetic defects,  
which act as the pair-breakers on both the $d$-wave and the $s$-wave states, were 
probably present in these studies.  

It is noteworthy that the effect of an abrupt rise in the specific heat jump 
at $T_c$ may be observed even in the purely $d$-wave superconductors in the 
presence of a perpendicular magnetic field (${\bf H}\parallel$ c-axis). 
The $s$-wave component in this case may be induced by the vortices. 
\cite{36}\\ 

We have derived the specific heat jump from a mean field weak-coupling theory, 
neglecting the fluctuations and the strong-coupling effects. As the observation 
of thermodynamic fluctuations in the specific heat of 
crystals of YBCO has been reported, \cite{25} we expect our BCS result to 
be modified by the deviations from the mean field approximation. We hope, 
however, that the feature of a sharp upturn in the specific heat will be still 
present. The strong-coupling corrections will rescale the scattering rates 
\cite{16} and may change the magnitude of the specific heat jump. \cite{26,30}\\

In conclusion, we have calculated the electronic specific heat difference 
between superconducting and normal state at the phase transition as a function 
of nonmagnetic impurity scattering rate in the general case of an anisotropic 
superconductor. We have found that the result depends on the symmetry 
of the order parameter, given by a function $e({\bf k})$, and that of the 
Fermi surface through the following FS averages: 
$\left<e\right>$, $\left<e^3\right>$ and $\left<e^4\right>$. 
A remarkably different dependence of the specific heat jump on the impurity 
concentration for the systems with $\left<e\right>=0$ and 
$\left<e\right>\neq 0$ is observed. 
We suggest that this effect may be used as a test for the $s$-wave component 
in the order parameter of the cuprates.\\

This work was supported by the Natural Sciences and Engineering 
Research Council of Canada.

\newpage
\begin{table}
\begin{center}
\caption{The elliptical Fermi surface averages of the powers of the normalized 
order parameter $e\left({\bf k}\right)=\left(k^{2}_{x}-k^{2}_{y}\right)
\left<\left(k^{2}_{x}-k^{2}_{y}\right)^{2}\right>^{-1/2}$.} 
\vspace*{0.2cm}
\begin{tabular}{||c|c||}
$\D \nu\left(\frac{c_x}{c_y}\right)$ & $\D\left[\frac{3}{2}-\frac{c_x}{c_y}
+\frac{3}{2}\left[\frac{c_x}{c_y}\right]^{2}\right]^{-1/2}$\\   
&\\
$\left<e\right>$ & $\D \nu\left(\frac{c_x}{c_y}\right)
\left[1-\frac{c_x}{c_y}\right]$\\  
&\\
$\left<e^2\right>$ & $ 1 $\\  
&\\
$\left<e^{3}\right>$ & $\D \nu^{3}\left(\frac{c_x}{c_y}\right)
\left[\frac{5}{2}\left[1-\left[\frac{c_x}{c_y}\right]^{3}\right]
-\frac{3}{2}\frac{c_x}{c_y}\left[1-\frac{c_x}{c_y}\right]\right]$\\ 
&\\
$\left<e^{4}\right>$ & $\D 16\nu^{4}\left(\frac{c_x}{c_y}\right)
\left[\frac{35}{128}\left[1+\frac{c_x}{c_y}\right]^{4}
-\frac{5}{4}\left[1+\frac{c_x}{c_y}\right]^{3}
+\frac{9}{4}\left[1+\frac{c_x}{c_y}\right]^{2}
-2\left[1+\frac{c_x}{c_y}\right]+1\right]$ 
\end{tabular}
\end{center}
\end{table}

\newpage
\section*{Figure Captions}
\noindent
Fig. 1. Jump in specific heat at $T_c$ normalized by the normal state specific  
heat at $T_c$ as a function of the normalized impurity scattering rate 
for $c_x/c_y=1$ i.e. $\left<e\right>=0$ (solid), 
$c_x/c_y=0.9$ i.e. $\left<e\right>\simeq 0.0742$ (short-dashed), 
$c_x/c_y=0.85$ i.e. $\left<e\right>\simeq 0.1139$ (dot-dashed), 
$c_x/c_y=0.8$ i.e. $\left<e\right>\simeq 0.1552$ (long-dashed):   
(a) Born limit, (b) unitary limit.\\ 

\noindent 
Fig. 2. Jump in specific heat at $T_c$ normalized by the normal state specific heat 
at $T_c$ as a function of the normalized critical temperature $T_c/T_{c_{0}}$ 
for $\left<e\right>=0$ (solid), $\left<e\right>\simeq 0.0742$ (short-dashed),
$\left<e\right>\simeq 0.1139$ (dot-dashed), $\left<e\right>\simeq 0.1552$
(long-dashed): (a) Born limit, (b) unitary limit. The insets show    
$\Delta C\left(T_c\right)/C_N\left(T_c\right)$ in the range of small $T_c$.\\ 

\noindent
Fig. 3. Normalized critical temperature $T_c/T_{c_{0}}$ as a function of 
the normalized impurity scattering rate for $\left<e\right>=0$ (solid), 
$\left<e\right>\simeq 0.0742$ (short-dashed), $\left<e\right>\simeq 0.1139$ 
(dot-dashed), $\left<e\right>\simeq 0.1552$ (long-dashed).\\ 

\noindent 
Fig. 4. Jump in specific heat at $T_c$ normalized by the normal state specific  
heat at $T_c$ as a function of impurity scattering rate 
for $\left<e\right>\simeq 0.6058$ ($c_x/c_y=0.3$) in the Born (dashed) 
and unitary (solid) limits.\\  

\noindent
Fig. 5. Position of the minimum in the normalized specific heat jump 
$\Delta C\left(T_c\right)/C_N\left(T_c\right)$ on $\varrho_c T_c/T_{c_{0}}$ 
axis (Figs. 1a, b) 
as a function of the $s$-wave component content $\left<e\right>$ 
in the Born (dashed) and unitary (solid) limits.\\  

\noindent
Fig. 6. Position of the minimum in the normalized specific heat jump
$\Delta C\left(T_c\right)/C_N\left(T_c\right)$ on $T_c/T_{c_{0}}$
axis (Figs. 2a, b)
as a function of the $s$-wave component content $\left<e\right>$ 
in the Born (dashed) and unitary (solid) limits.


\begin{references}
\bibitem{29} Since the average value over FS may be interpreted as the inner 
product, it yields from Cauchy-Schwartz inequality and the normalization 
condition $\left<e^2\right>=1$ that $\left<e^{4}\right>\ge 1$. Therefore another 
limiting case $\left(\Delta C\left(T_c\right)/C_{N}\left(T_c\right)\right) 
_{\varrho_c\rightarrow\infty}=1$ requiring $\left<e\right>=0$ and 
$\left<e^{4}\right>=2/3$ must be excluded.
\bibitem{27} A convincing argumentation for a "gyroscopic" behavior of 
the order parameter which does not change its orientation accross the twin 
boundaries and therefore is not dependent on the orthorhombic directions 
in the crystal was presented in Ref. 1.
\bibitem{30} It has been shown in Ref. 26 that in a phonon mediated isotropic  
superconductor, $\Delta C(T_c)/\gamma (0)T_c$,  
where $\gamma (0)$ is the zero temperature Sommerfeld constant, 
can attain values about $2.5$ times larger than the BCS value of $1.426$.
\end{references}
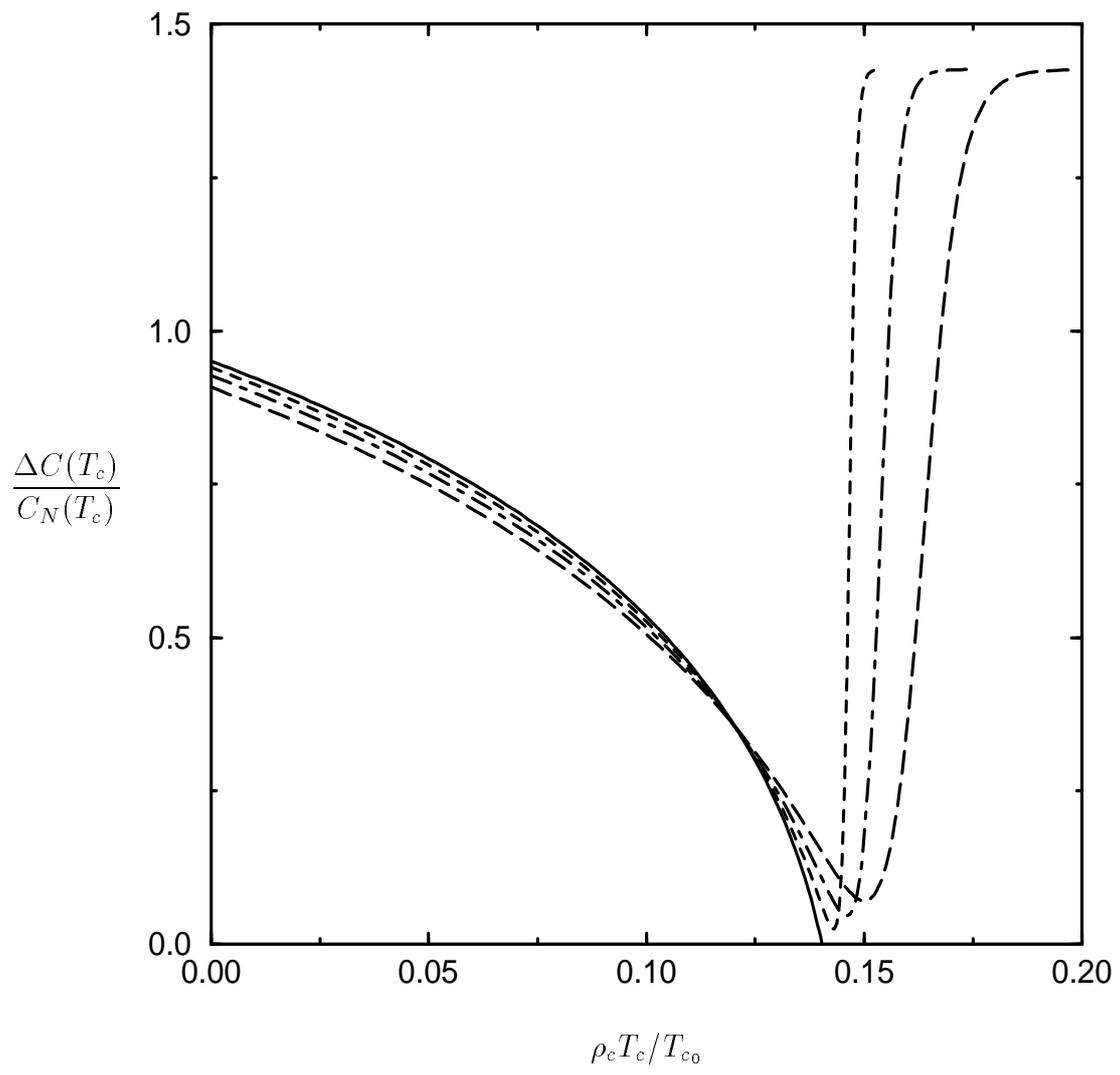
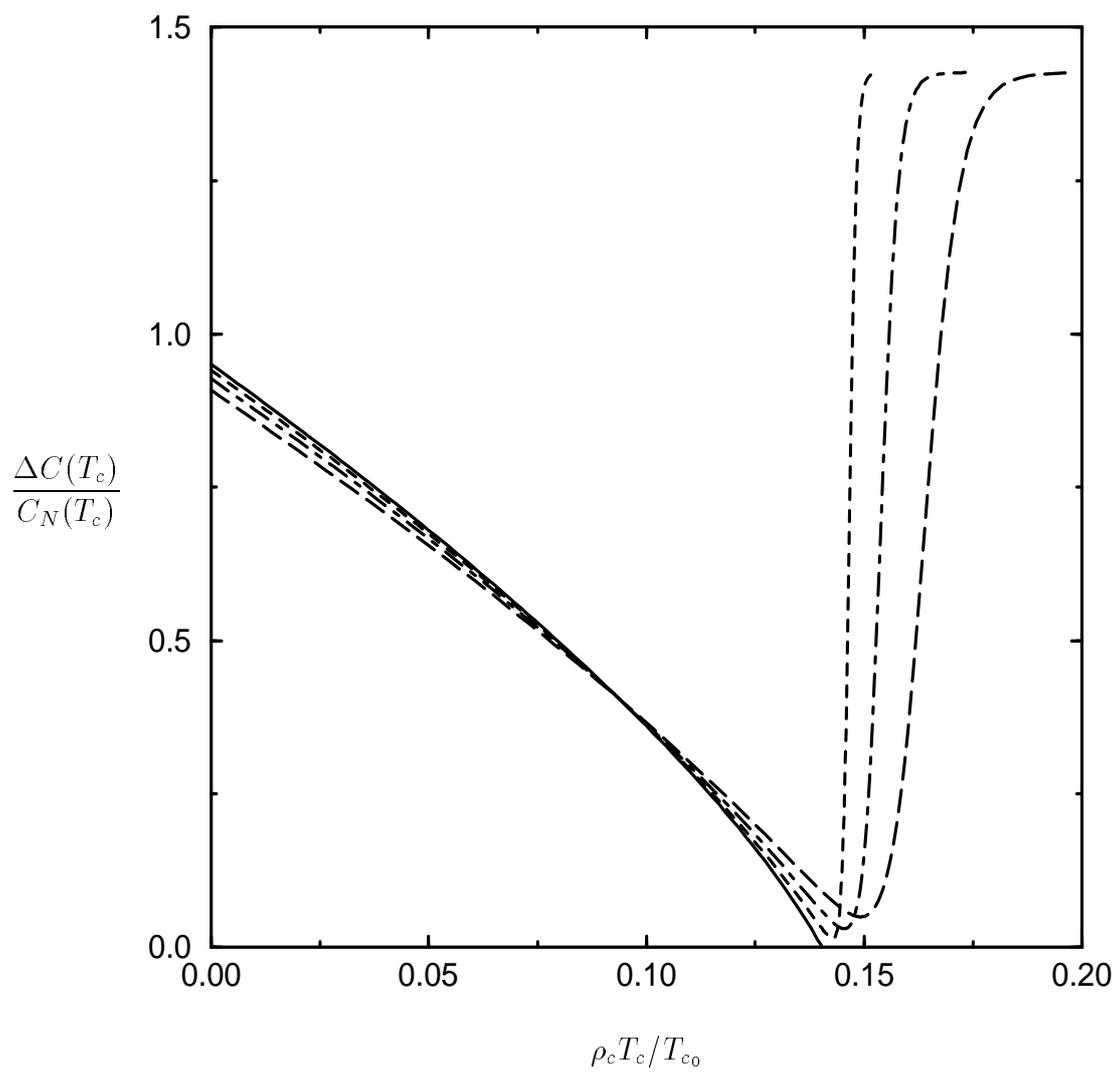
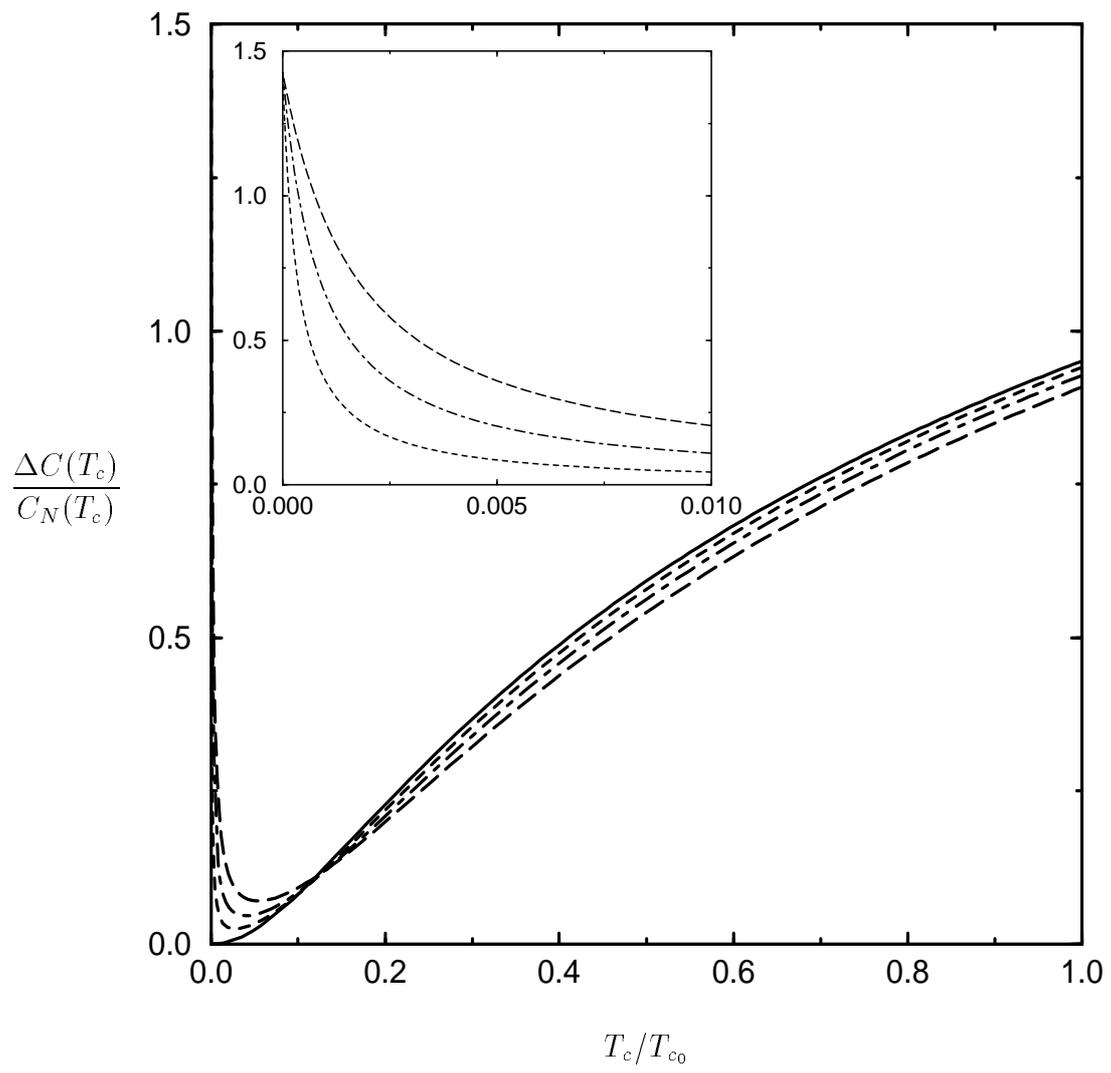
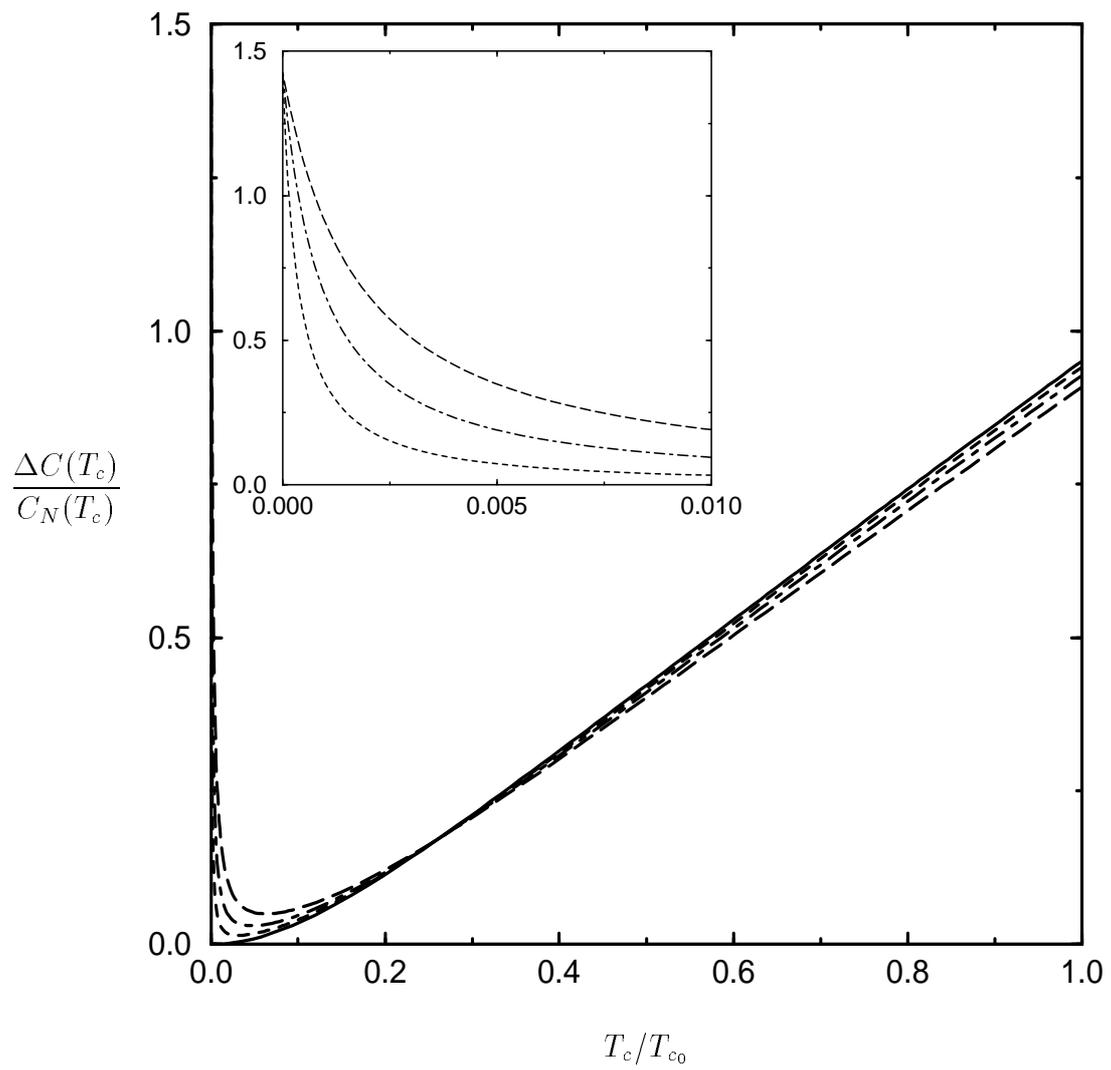
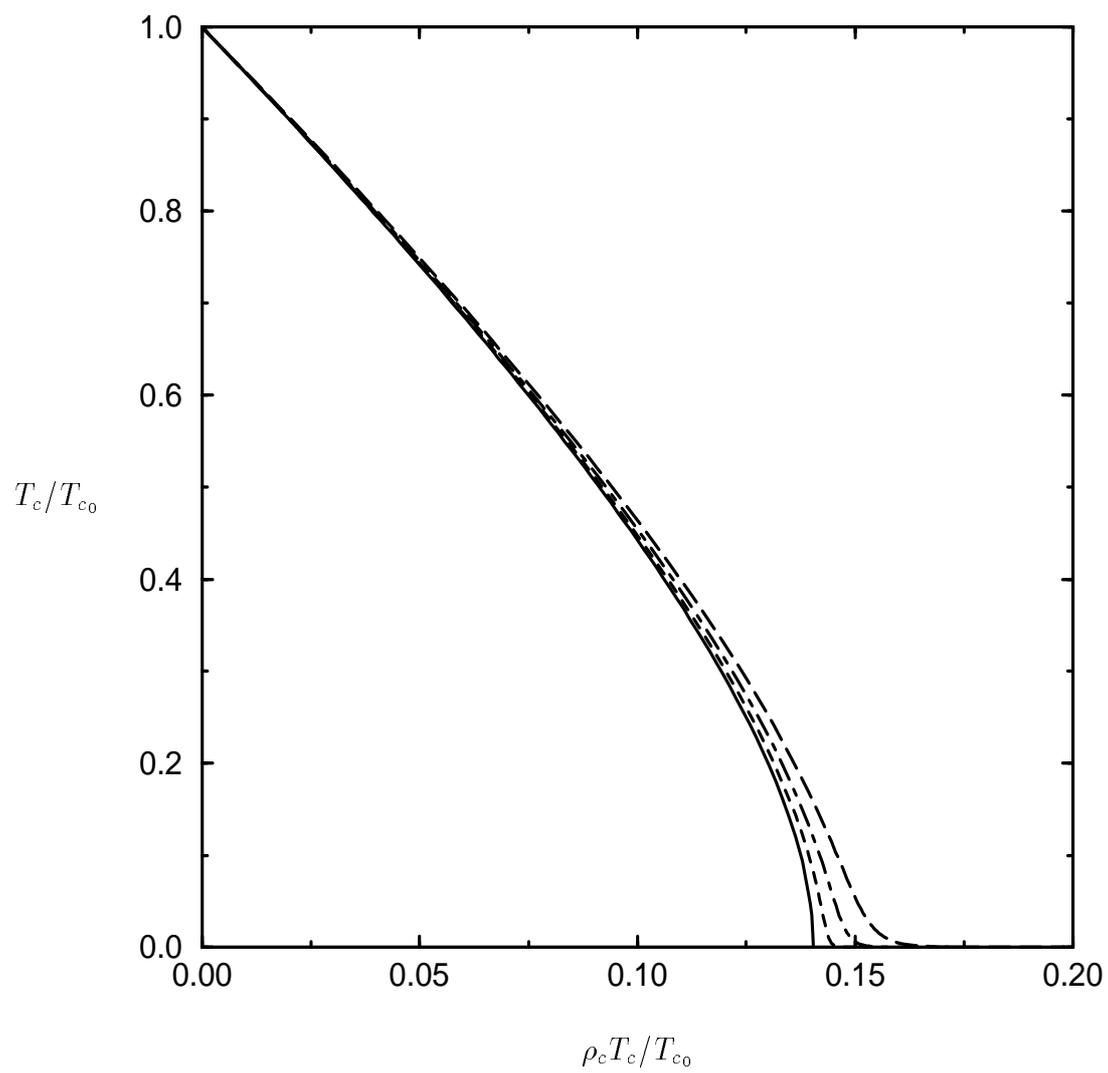
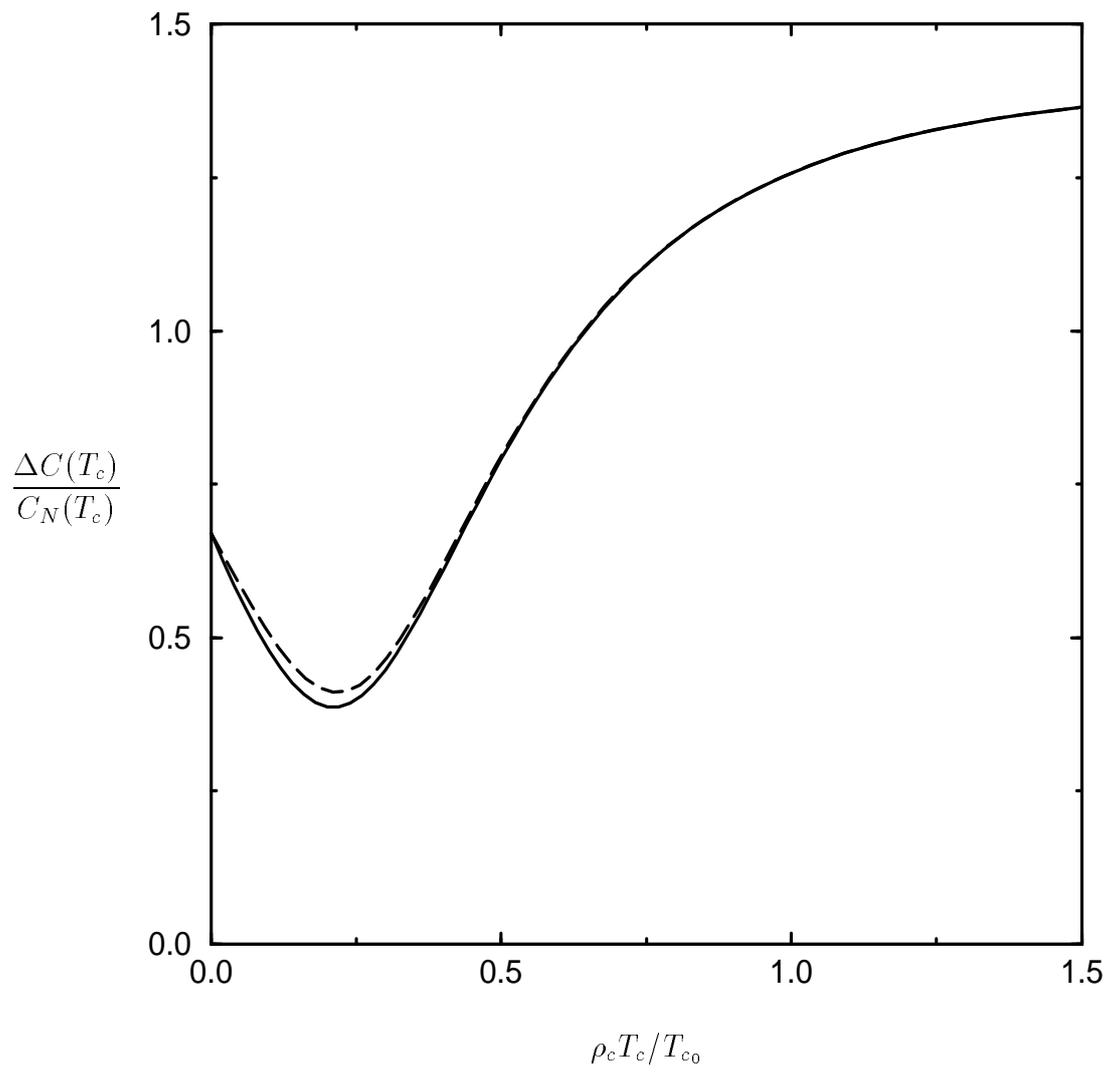
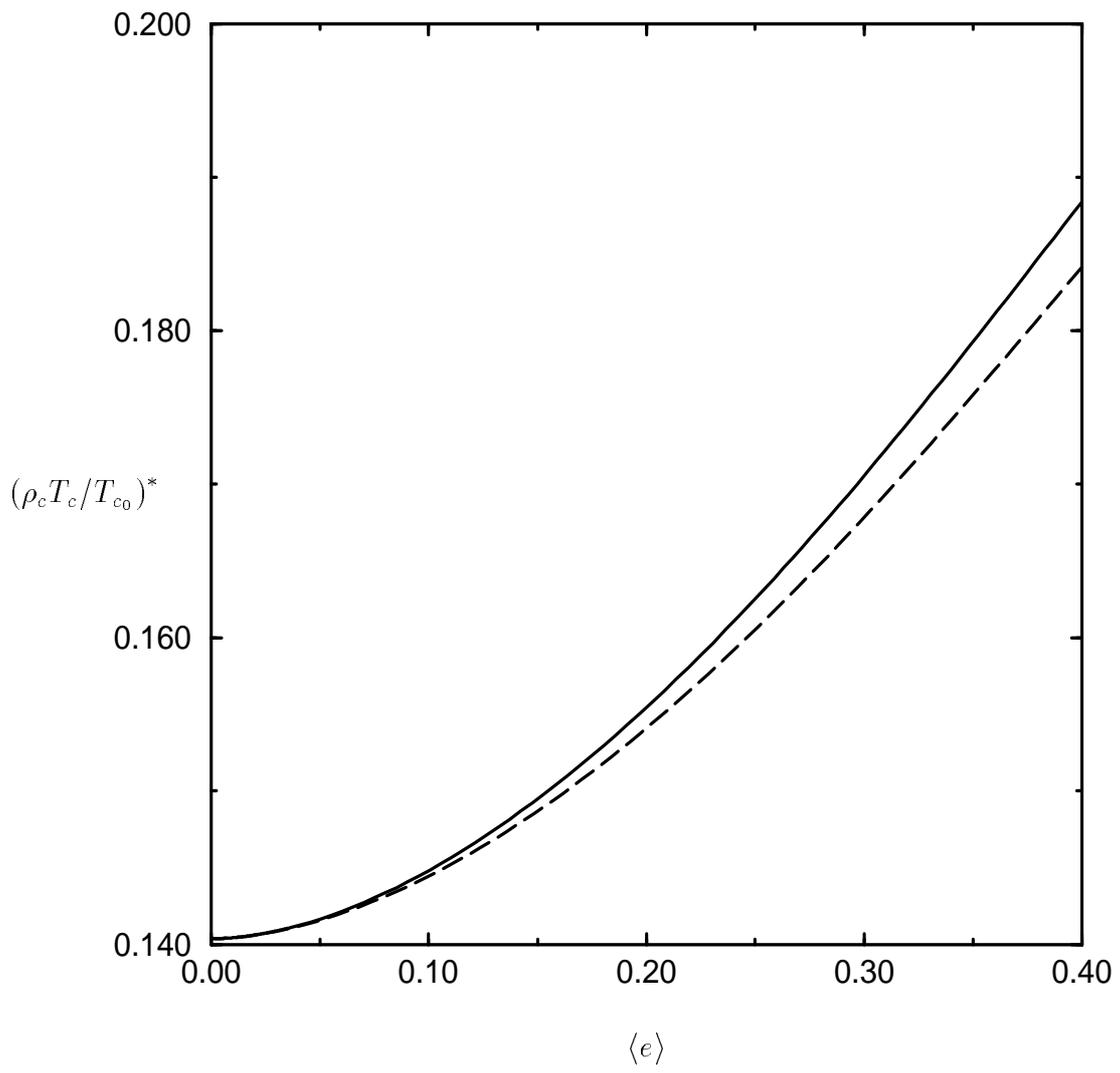
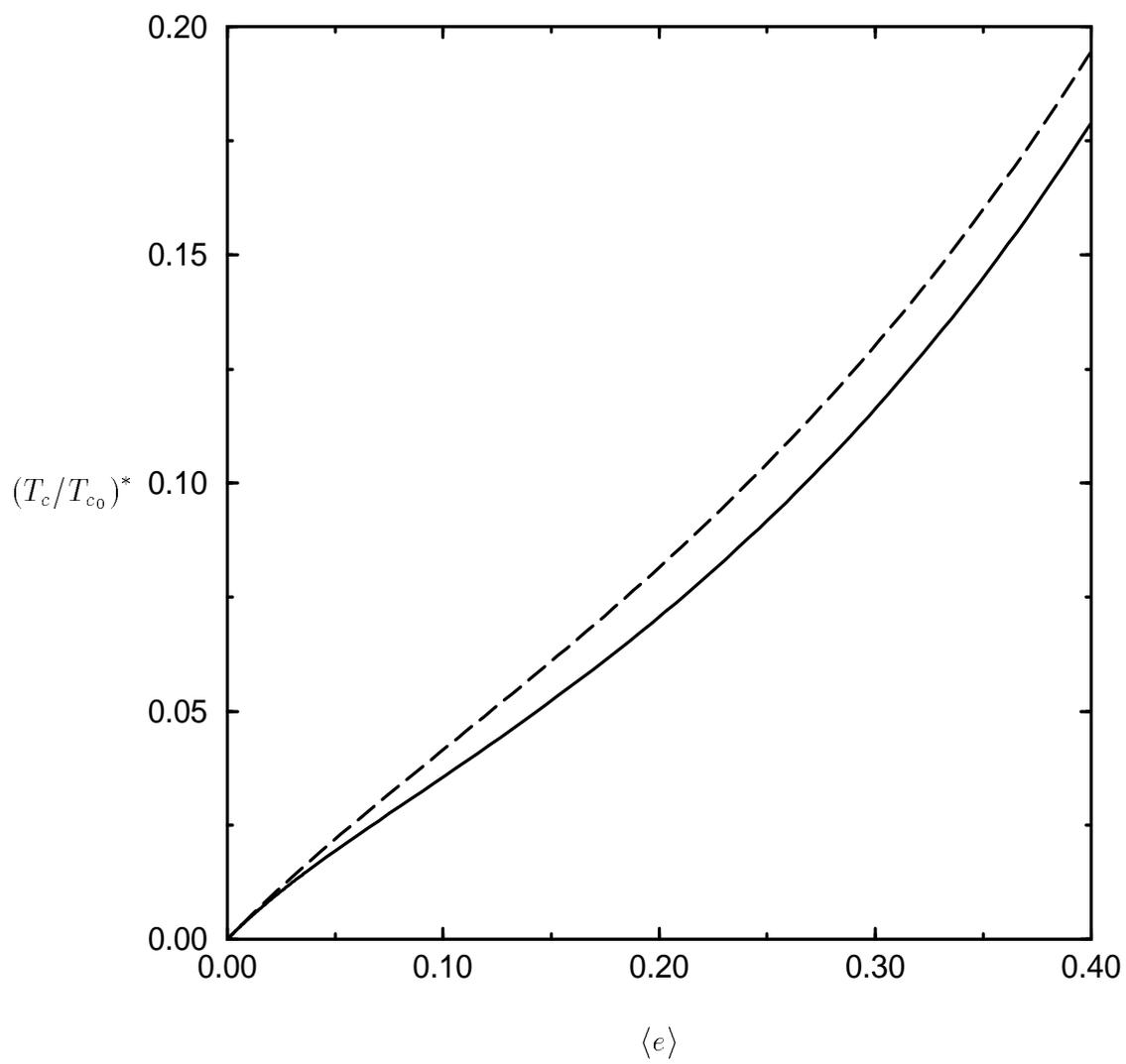
\end{document}